\documentclass[aip,apl,reprint]{revtex4-1}

\usepackage[compact]{titlesec}
\usepackage{graphicx}
\usepackage{sidecap}
\usepackage{float}
\usepackage{calc}
\usepackage{pifont}
\usepackage{amssymb}
\usepackage{epsfig}
\usepackage{amsmath}
\usepackage{color}
\usepackage{braket}

\sidecaptionvpos{figure}{c}

\DeclareMathAlphabet{\mathitb}{OT1}{cmr}{bx}{sl}
\begin{document}

\renewcommand{\thefootnote}{\fnsymbol{footnote}}

\title{Electrode Design for Antiparallel Magnetization Alignment in Nanogap Devices}

\author{G. D. Scott}
\email{gavin.scott@nokia.com}
\affiliation{
Bell Laboratories, Nokia, 600 Mountain Ave, Murray Hill, NJ 07974}

\date{\today}

 \begin{abstract}

The ability to manipulate the relative magnetization alignment between ferromagnetic source and drain electrodes attached to a molecule or small quantum dot is a prerequisite for a number of spintronic device applications.  The influence of electrode shape and field orientation on pair-wise magnetization reversal mechanisms in nanogap and point-contact structures is investigated here using micromagnetic simulations.  A favorable device geometry and setup are identified for enabling planar, monodomain source and drain electrodes with a magnetization alignment that may be controllably switched between a parallel and anti-parallel configuration.

\end{abstract}

\maketitle

\section{INTRODUCTION}
\vspace{2mm}
Understanding and controlling the magnetization of micro- and nano-scale structures has been an essential component of spintronics research for purposes such as data storage, information processing, and magnetic field sensing.\cite{Bader2006,Cowburn2000}  Many studies have focused on the magnetization reversal processes in both isolated and ordered arrays of magnetic nanoelements,\cite{Cowburn2000,Lai2002,Novosad2003,Vavassori2004,Carace2006,Lau2007,Yin2011} but there are a number of novel applications which require control of the magnetization alignment between isolated pairs of planar electrodes attached to a molecule or quantum dot.  These include spin state detection of individual magnetic atoms or molecules,\cite{Chen2008} nanoelectromechanical single electron shuttle devices,\cite{Kulinich2014,Ilinskaya2015} and tools for the investigation of quantum critical phenomena relevant to heavy fermion materials.\cite{Kirchner2005}

Reliable methods have been established to control the relative orientation of magnetization between stacked magnetic layers separated by non-magnetic material, as in spin-valve devices.\cite{Parkin2003}  On the other hand, controlling the magnetization orientation between two adjacent thin film structures with nanometer-scale separation has proven to be more difficult.  Experimental efforts with break junction devices have found limited success by producing what are likely to be in-plane flux-closure (vortex) magnetization distributions.  Here, micromagnetic modeling is used to analyze the impact of geometry and magnetic field orientation on the magnetization reversal mechanisms in closely spaced source-drain electrode-pairs.  The goal is to find a realistic design and device setup for obtaining electrodes with nanometer separation and uniform, monodomain magnetization distributions that may be reliably switched between two states of relative alignment:  parallel (P) and anti-parallel (AP).\\

\section{NUMERICAL SIMULATION}
\vspace{2mm}
The magnetization reversal process is investigated here with the finite-element micromagnetic simulation package, $Nmag$.\cite{Nmag}  The source and drain electrodes are initialized into a P configuration with an in-plane external magnetic field, $\textbf{H}_{ext}$, roughly equal to $\textbf{H}_{s}$, the field required to saturate the magnetization in a particular direction.  A small field component ($1\%$) orthogonal to $\textbf{H}_{ext}$ is added to break the symmetry.  The Landau-Lifshitz-Gilbert equation is integrated numerically over time\cite{Nmag,Wei2007} at incrementally changing values of external field as $\textbf{H}_{ext}$ is swept from $\textbf{H}_{s}$ to -$\textbf{H}_{s}$ and then back to $\textbf{H}_{s}$.  Permalloy (Ni$_{80}$Fe$_{20}$) was chosen as a test material due to its lack of magnetocrystalline anisotropy and common usage among research and industrial applications.  The material parameters used were saturation magnetization $M_{s} = 7.958 \times 10^5~A/m$, exchange coupling strength $C = 13 \times 10^{-12}~J/m$, and damping parameter $\alpha = 0.5$ or a more realistic value of $\alpha = 0.05$ in selected cases.

\begin{SCfigure*}
  \centering
  \includegraphics[height=4.6cm]{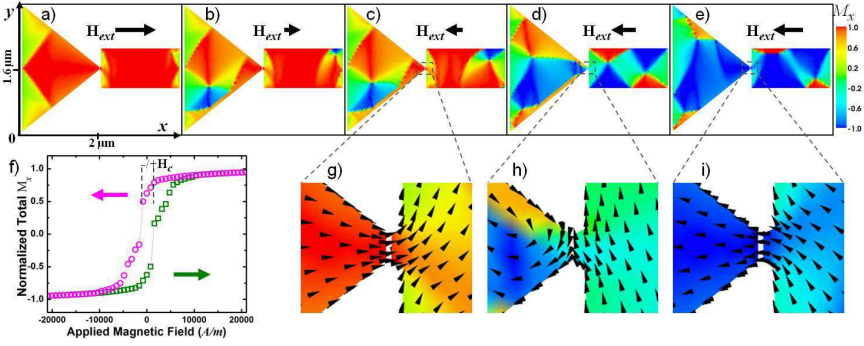}
  \caption{Snapshots of simulated magnetization reversal in the T-R geometry as $\textbf{H}_{ext}$ is reduced from positive saturation (a), at remanence (b), immediately before (c) and after (d) the jump at $\textbf{H}_{c}$, and further past the jump (e). Colorbar indicates normalized $\emph{x}$-component of magnetization, M$_\emph{x}$.  (f) Total M$_\emph{x}$ $\emph{vs.}$ $\textbf{H}_{ext}$ calculated as $\textbf{H}_{ext}$ is swept down from $\textbf{H}_{s}$ (magenta circles) and back up (green squares).  Arrows indicate sweep direction.  (g)-(i) Close up of nanogap region corresponding to (c)-(e), respectively.  Cones denote local magnetization direction.}
  \label{TriRect}
\end{SCfigure*}

Earlier experimental efforts to create magnetic nanocontact devices that could be alternated between P and AP configurations focused on shape anisotropy, utilizing two electrodes with different shapes,\cite{Pasupathy2004,Keane2006,Yoshida2013} like the triangle-rectangle (T-R) geometry in Fig.~\ref{TriRect}, connected via atomic-scale contacts or tunneling gaps formed with standard break junction techniques.  Shape anisotropy accounts for the relationship between the mean magnetization direction and the geometrical form of a magnetic element leading to a demagnetizing field magnitude dependent upon the direction of an applied external field.  It was presumed that as $\textbf{H}_{ext}$ was swept from $\textbf{H}_{s}$ through $\textbf{H}_{ext} = 0~A/m$, the different coercivities associated with the mismatched geometries would enable the magnetization of one electrode to reverse polarity prior to the other, leading to an AP magnetization alignment.  However, this premise ignores the magnetostatic coupling between the two electrodes, which becomes increasingly relevant as the inter-electrode separation decreases.\cite{Novosad2003}\\

\section{RESULTS AND DISCUSSION}
\vspace{2mm}
The T-R model shown in Fig.~\ref{TriRect} was designed with $20~nm$ thickness, and length scales that approximated the dimensions used in Refs.~\citenum{Pasupathy2004,Keane2006,Yoshida2013}.  Results of the simulations with this model validate that it will not accurately produce a uniform AP configuration between source and drain contacts.  The magnetization reversal may be characterized here in part with snapshots of the longitudinal (\emph{x}-axis) magnetization component magnitudes, denoted by colormap, and the transient spin configuration, denoted by vector field, acquired at incrementally changing values of $\textbf{H}_{ext}$.  As $\textbf{H}_{ext}$ is reduced from saturation in the +$\emph{x}$ direction a vortex core nucleates near an edge in each lead and propagates roughly along the ${y}$-axis (Fig.~\ref{TriRect}(a)-(e)).  The reversal process is further delineated with a graph of the sum of longitudinal magnetization vector components plotted as a function of $\textbf{H}_{ext}$ (Fig.~\ref{TriRect}(f)).  This produces a hysteresis loop as the magnetization distribution within the electrodes undergoes an abrupt shift at some critical field, $\textbf{H}_{c}$.  An AP configuration would be expected at values of $\textbf{H}_{ext}$ for which $\Sigma{M_x}\rightarrow0$.  When the field is swept past $-\textbf{H}_{c}$ two vortices with opposite chirality exist in each lead.  As $\textbf{H}_{ext}$ is further reduced the vortices move to the edges and are ultimately annihilated.  The evolving spin configuration in the T-R geometry (Fig.~\ref{TriRect}(g)-(i)) enables the magnetization of the electrodes to transition from a P state to a non-collinear configuration in which neighboring vortex states may effectively lead to a local AP magnetization orientation between the domains immediately on the left and right sides of the tunneling gap.  Here the remainder of the electrodes are in a mixed state of buckling modes and vortex-propagation modes, as is common for the magnetization reversal in ferromagnetic thin film structures.\cite{Wei2007}  Simulations were also run on other T-R pairs with various scales, length/width ratios, and gap sizes and were found to produce qualitatively similar results (i.e. mixed state modes and the possibility of a local AP alignment at the break junction site).

Magnetization reversals have been studied with variously shaped elements, but ellipses have received particularly close attention partially due to their well defined uniaxial shape anisotropy.  The long (short) diameter of an ellipse corresponds to its easy (hard) axis, along which the demagnetizing field and magnetostatic energy are minimized (maximized).  Magnetization reversals with closely-spaced ellipses rely on their magnetostatic dipole-dipole interaction in addition to the shape anisotropy of the individual elements.  In a design used in Refs.~\citenum{Bolotin2006,Shi2007} a pair of nominally identical elliptical contacts are utilized as source and drain electrodes.  The long axis of the ellipses are aligned parallel to one another and perpendicular to both the bridge connecting them and the direction of current flow.  The external field, $\textbf{H}_{ext}$, is aligned parallel to the long axis of the ellipses.  

\begin{figure}[b!]
\begin{center}
\vspace{-7mm}
\includegraphics[scale = .255]{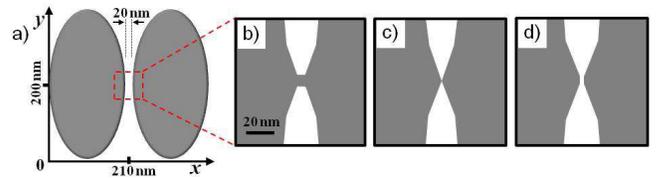}
\end{center}
\vspace{-6mm}
\caption{(a) Example of ellipse-pair geometry with the \emph{ideal} model version.  Enlarged view of constriction region, corresponding to red dashed box in (a), highlights variations between the (b) \emph{unbroken}, (c) \emph{point-contact}, and (d) \emph{nanogap} variations of the model.
}
\label{Geo}
\end{figure}

\begin{figure}[t!]
\begin{center}
\includegraphics[scale = .55]{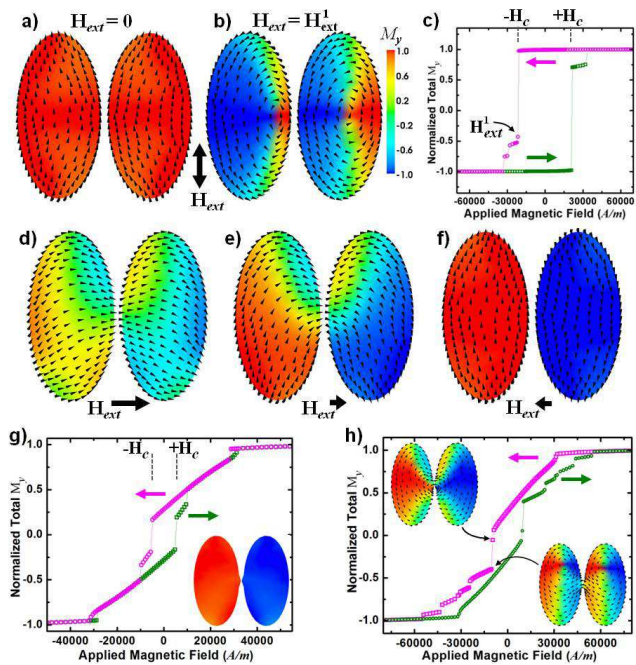}
\end{center}
\vspace{-5mm}
\caption{Snapshots of simulated magnetization distribution using the \emph{ideal} ellipse-pair geometry with a transverse ($\emph{y}$-axis) field orientation at remanence (a), and at $\textbf{H}^{1}_{ext}$ (b) immediately after the jump at -$\textbf{H}_{c}$ as labeled in (c).  Colorbar (same for all of Fig.~\ref{Ellipses}) indicates local $M_y$ magnitude and black cones denote local magnetization direction.  Intermediate states lead to vortices in both ellipses, but the AP configuration with uniform magnetization is never achieved.  (c) Calculated hysteresis loop of normalized $M_y$ $\emph{vs.}$ $\textbf{H}_{ext}$, from same simulation as data in (a) and (b).  (d) Snapshots using the same model with a longitudinal ($\emph{x}$-axis) field orientation as $\textbf{H}_{ext}$ is initially reduced from saturation (d), shortly past remanence (e), and immediately after the jump at -$\textbf{H}_{c}$ (f).  (g) Calculated hysteresis loop of $M_y \emph{vs.} \textbf{H}_{ext}$, using the $\emph{nanogap}$ ($3~nm$ gap) device variation.  Inset: Monodomain-like AP magnetization alignment still possible when a gapped constriction is included in the simulation geometry.  (h) Same as (g), but for the $\emph{point-contact}$ model.  Insets: top left and bottom right show the transient spin configuration before and after, respectively, the jump at -$\textbf{H}_{c}$.  The uniform AP configuration is never achieved.
}
\label{Ellipses}
\end{figure}

To simulate the evolving magnetization distribution in devices akin to those in Refs.~\citenum{Bolotin2006,Shi2007}, we used a $15~nm$ thick base model with lateral dimensions as shown in Fig.~\ref{Geo}(a).  Simulations were alternately performed using four basic design alternatives (Figs.~\ref{Geo}(b)-(d)), each tested with $\textbf{H}_{ext}$ aligned along the $\emph{x}$-axis and $\emph{y}$-axis.  The \emph{unbroken} version consists of the two ellipses connected with the a small constriction, or bridge, of finite width ($4-12~nm$).  The \emph{point-contact} has a continuous bridge constricting down to a single point.  The \emph{nanogap} includes a tapered bridge possessing a gap between the left and right ends of the constriction from $6~nm$ down to $2~nm$ - an approximate lower bound on the length scale, below which the validity of the calculations may be ambiguous.  Lastly, the \emph{ideal} set of leads consists of the ellipses only (i.e. no constriction).  This allowed for the effects of shape anisotropy and dipole-dipole interaction on the ellipses to be deduced both with and without the influence of the extraneous bridge material.

Simulations using an easy axis field orientation for both the \emph{ideal} and \emph{nanogap} versions of this device model demonstrate an abrupt switch between a P state in which the magnetization of both electrodes points almost entirely in the same direction (+$\emph{y}$) to an intermediate configuration with both ellipses exhibiting displaced single vortex states (Fig.~\ref{Ellipses}(a),(b)).  This results in a local AP configuration between the closest points on the ellipses for the \emph{ideal} model and on opposing sides of the tunneling barrier in the \emph{nanogap} model.  The colorbar indicates the magnitude of the transverse magnetization component and the cones indicate local magnetization orientation.  When the \emph{point-contact} and \emph{unbroken} versions of this device model are used in the simulation, comparable intermediate states are not reached.  Instead the abrupt transition at the coercive field takes the system from a P state in which the magnetization of both electrodes is nearly uniform in the +$\emph{y}$-direction to the same P state in the -$\emph{y}$-direction.

A distinctly different reversal process is demonstrated by simulations performed using the same electrode geometry and finite element mesh, but with $\textbf{H}_{ext}$ applied along the hard axis. Representative snapshots of the simulated magnetization distribution using the \emph{ideal} device model with a longitudinal field orientation are presented in Fig.~\ref{Ellipses}(d)-(f).  For all versions of this device model, as $\textbf{H}_{ext}$ is reduced from $\textbf{H}_{s}$ the magnetization in the regions along the outer edges of the ellipses begins to rotate in order to reduce the magnetostatic energy.  In the example shown in Fig.~\ref{Ellipses}(d), the magnetization of the left electrode turns upward and the magnetization of the right electrode turns downward.  At the same time the constriction region (when present) remains homogenously magnetized.  As the field is further reduced, a buckling-type configuration arises which encompasses both ellipses such that they exhibit a joint magnetization distribution pattern with an inverted U-shape (Fig.~\ref{Ellipses}(e)).  For the \emph{ideal} and \emph{nanogap} models, this distribution pattern persists as $\textbf{H}_{ext}$ is swept past zero until an abrupt re-distribution occurs at the critical field, resulting in the configuration shown in Fig.~\ref{Ellipses}(f) and Fig.~\ref{Ellipses}(g) inset, respectively.  Here the two electrodes exhibit nearly homogenous magnetic moments with opposite polarity, which is precisely the target magnetization configuration.  For the \emph{unbroken} and \emph{point-contact} models this AP state is never reached, thus the presence of a tunneling gap is critical, consistent with the fact that the exchange energy would oppose a large rotation of magnetic moments between two adjacent atoms in a chain.  The magnetization distributions closest to the AP configuration can be seen in the insets of Fig.~\ref{Ellipses}(h).  Further simulations examining the tolerance on the field angle for obtaining the AP configuration showed that this intermediate state is extremely robust with respect to an out-of-plane component of $\textbf{H}_{ext}$, but is quite sensitive to the inclusion of a \emph{y}-axis component, such that it may fail to form for in-plane field angles deviating beyond a small range ($\sim2^{\circ}$) from the \emph{x}-axis.

Analogous simulations were also run with different variations of the ellipse-pair base model.  These included versions of the above geometry with uniform in-plane scaling ($\times0.75$, $\times1.2$ and $\times1.5$), modified ellipse eccentricity (height/width ratios of $1.5$, $2.0$, and $2.5$), and increased material thickness ($20~nm$, $25~nm$, and $30~nm$) and ellipse spacing ($30~nm$ and $40~nm$).  Every geometrical alteration affects the magnetostatic energy of a device and can impact its magnetic response to an external field.  In general, smaller, thinner ellipses with closer spacing and larger eccentricity exhibited qualitatively similar reversal processes, but with increased ranges of $\textbf{H}_{ext}$ over which the monodomain-like AP configuration was maintained, which may be a beneficial attribute.  However, advantages of reduced dimensions will be constrained as the electrodes become more difficult to create and contact with existing fabrication techniques.  The ellipse-pair geometry in Fig.~\ref{Ellipses} presents a practical solution (i.e. easily within fabrication limitations) with a magnetization reversal process representative of the various models tested, but it does not preclude a more optimal design for a given application.

Another variation was designed to more closely mimic the dimensions of the permalloy ellipses used in Refs.~\citenum{Bolotin2006,Shi2007}, which were deposited partially on top of pre-existing gold contacts to avoid the formation of an oxide barrier, thereby creating a step near the middle of each ellipse.  Simulations performed using this non-planar geometry indicate that for the hard axis field orientation, the magnetization distribution does not exhibit the jump to the AP configuration.  Instead a transformation occurs resulting in single vortex states in each ellipse, much like the intermediate state of planar ellipses with an easy axis field orientation shown in Fig~\ref{Ellipses}(b).  For an applied field along the easy axis of the non-planar model a mixed state is reached in which one ellipse is homogeneously magnetized along the $\emph{y}$-axis while the opposing electrode contains a single displaced vortex.\\

\section{CONCLUSION}
\vspace{2mm}
The numerical simulations performed here indicate that the ellipse-pair electrode design is a favorable geometry for obtaining the desired AP magnetization configuration in break junction devices.  When an external magnetic field is oriented along the hard axis of ellipses with nanometer-scale separation, the system may be controllably tuned from a P configuration to a state in which the source and drain contacts exhibit monodomain-like magnetization distributions with AP alignment, whereas an easy axis field orientation will only result in vortex states within each ellipse.  Furthermore, we find that the magnetization distribution will be influenced by the non-planar shape produced as a result of a fabrication process that involves partially overlapping the magnetic elements on top of other thin films of comparable thickness.  Overcoming fabrication obstacles may provide further insights into the magnetization reversal mechanisms in the structures addressed here by allowing for a direct comparison between simulated and experimental data.  For instance, measurements of magnetoresistance in break junction devices with the geometries described above could be examined to substantiate the accuracy of these numerical results.

\vspace{-1mm}
\bibliographystyle{apsrev}
			
\bibliography{PyMdesA_noURL}

\end{document}